\documentclass[11pt,a4paper]{article}

\usepackage{graphicx}
\usepackage{amsmath}
\usepackage{amssymb}
\usepackage{palatino}
\usepackage{mathpazo}
\usepackage{caption}
\usepackage[norounding,point]{rccol}
\newcommand{\ud}{\,\mathrm{d}}

\begin{document}

\thispagestyle{empty}

\setlength{\hoffset}{0mm}
\setlength{\parindent}{0pt}
\setlength{\parskip}{1ex plus 0.5ex minus 0.2ex}
\setlength{\marginparwidth}{0pt}
\setlength{\oddsidemargin}{0pt}
\setlength{\evensidemargin}{0pt}

%\fontsize{11}{12}\selectfont

\title{Correction of beam-beam effects in luminosity measurement at ILC} 

\author{S. Luki\'c and I. Smiljani\'c\\
		Vin\v{c}a Institute of Nuclear Sciences, University of Belgrade, Serbia}
\date{\today}

\maketitle 

\begin{abstract}
Three methods for handling beam-beam effects in luminosity measurement at ILC are tested and evaluated in this work. The first method represents an optimization of the LEP-type asymmetric selection cuts that reduce the counting biases. The second method uses the experimentally reconstructed shape of the $\sqrt{s'}$ spectrum to determine the Beamstrahlung component of the bias. The last, recently proposed, \emph{collision-frame} method relies on the reconstruction of the collision-frame velocity to define the selection function in the collision frame both in experiment and in theory. Thus the luminosity expression is insensitive to the difference between the CM frame of the collision and the lab frame. 
%Both the second and the third method are accurate to better than one permille of the total luminosity, but 
The collision-frame method is independent of the knowledge of the beam parameters, and it allows an accurate reconstruction of the luminosity spectrum above 80\% of the nominal CM energy. However, it gives no precise infromation about luminosity below 80\% of the nominal CM energy. The compatibility of diverse selection cuts for background reduction with the collision-frame method is addressed.
\end{abstract}

\section{Introduction}

Luminosity, $L$, and luminosity spectrum, $\mathcal{L}(E_{CM})$, are key input to analyses of most measurements at a linear collider, including mass and cross-section measurements, as well as the production-threshold scans. Precision measurement of the luminosity is thus essential for the physics programme at a linear collider. The standard way to measure luminosity is to count Bhabha-scattering events recognized by coincident detection of showers in the fiducial volume (FV) in both halves of the luminometer in the very forward region in a given energy range. The number of events $N$ is then divided by the Bhabha cross section $\sigma$ integrated in the corresponding region of the phase space. This can be formally expressed as follows,

\begin{equation}
\label{eq-luminosity}
L = \frac{N(\Xi(\Omega^{lab}_{1,2}, E^{lab}_{1,2}))}{\sigma(Z(\Omega^{CM}_{1,2}, E^{CM}_{1,2}))},
\end{equation}

Here $\Xi(\Omega^{lab}_{1,2}, E^{lab}_{1,2})$ is a function describing the selection criteria for counting the detected events based on the angles $\Omega^{lab}_{1,2}$ and energies $E^{lab}_{1,2}$ of the final particles in the lab frame, and $Z(\Omega^{CM}_{1,2}, E^{CM}_{1,2})$ is a function describing the corresponding region of phase space in the CM frame over which the cross section is integrated. These functions can be expressed as products $\Xi = \prod_i \xi_i$ and $Z = \prod_i \zeta_i$ where the functions $\xi_i$ and $\zeta_i$ are based on specific topological and kinematical properties of the detected/generated pair. For each $i$, the physical meaning of $\xi_i$ and $\zeta_i$ corresponds to each other, although their mathematical form may be different, as explained in Sec. \ref{sec-coll} and Ref. \cite{Luk12}. The set of functions $\xi_i$ and $\zeta_i$ includes the angular selection requiring both particles to be detected in the FV, as well as the energy range selection and possible further cuts to eliminate background.

Bhabha scattering is a well known QED process, for which cross-section calculations with relative uncertainty better than $10^{-3}$ are available \cite{Jad03}. However, one may note that $\Xi$ and $Z$ operate on kinematical arguments in different reference frames. The cross section is typically calculated by Monte Carlo integration using an event generator in the CM frame of the process. On the other hand, the experimentally detected Bhabha particles are affected by the beam-beam effects at bunch collision \cite{Yok91, Sch96}. Because of the random and asymmetric momentum loss when electrons emit Beamstrahlung, the CM frame of the Bhabha process is different for every colliding pair, and in general it does not coincide with the laboratory frame. As a consequence, the angles and energies of the outgoing particles seen in the laboratory frame are different than in the CM frame. Therefore, if the selection criteria are applied in the same form for the detected events as for the cross-section integration, different regions of the phase space will be covered, leading to a systematic bias in the luminosity measurement. In addition, the energy loss due to Beamstrahlung leads to a counting bias because of the energy cuts.

At LEP, uncertainties arising from the ISR and the Beamstrahlung emission were usually minimized by designing asymmetric selection cuts \cite{Opal00, Aleph00, L3_00}. Similar techniques were proposed for ILC as well, where the beam-beam effects are much more intense \cite{Stahl05, Rim07}. In this context, Rimbault et al. also proposed to use the experimentally reconstructed form of the $\sqrt{s'}$ spectrum to determine the remaining counting bias \cite{Rim07}. Two methods based on these two concepts will be presented here, and their effectivenes will be tested.

Finally, a recently proposed way of handling the beam-beam effects will be tested for ILC \cite{Luk12}. The guiding idea of the new method is to define $\Xi$ and $Z$ such that the counting rate is independent of the reference frame. This can be achieved by expressing some of the factors $\xi_i$ and $\zeta_i$ in terms of Lorentz-invariant kinematic parameters, and/or by designing $\xi_i$ and $\zeta_i$ in such a way that the reconstructed experimental count and the cross-section integration limits are defined in the same reference frame. If these conditions are met, the Bhabha count is bias-free, and requires no correction by beam-beam simulation, and no precise knowledge of the beam parameters. 

In Sec. \ref{sec-phys}, the physical processes affecting the luminosity measurement will be outlined, and the event simulation methods used in this work will be briefly described. In Sec. \ref{sec-corr}, the three methods of handling the beam-beam induced counting bias will be described and tested on simulated events. In Sec. \ref{sec-add-cuts}, the compatibility of the collision-frame method with various additional selection cuts of interest for background removal is discussed. In the conclusions, the performance of these methods for the final precision of the luminosity measurement will be summarized and discussed.

\section{Beam-beam effects on the Bhabha scattering and event simulation}
\label{sec-phys}

\subsection{Summary of the physical processes affecting the luminosity measurement}
\subsubsection*{Beamstrahlung}

Due to the pinch effect during the bunch collision, particles from both bunches may emit Beamstrahlung photons and so lose energy and momentum before the interaction. Thus, in general, the center-of-mass (CM) energy of an electron-positron pair is smaller than the nominal energy of the collider $E_{CM} < E_0 \equiv 2 E_{beam}$. The center-of-mass (CM) energy distribution at this stage is that of the actual luminosity spectrum $\mathcal{L}(E_{CM})$. The probability of the Bhabha scattering scales with $1/E^2_{CM}$, resulting in the CM energy distribution of the Bhabha events,

\begin{equation}
\label{eq-bhabha-spec}
 \mathcal{B}(E_{CM}) \propto \mathcal{L}(E_{CM})/E^2_{CM} 
\end{equation}

Beside that, due to the loss of momentum prior to the collision, the CM reference frame has a non-zero velocity with respect to the lab frame, and the detected particle polar angles $\theta^{lab}_1$ and $\theta^{lab}_2$ are boosted relative to the angles in the CM frame, which on average increases the acollinearity. In this way, Beamstrahlung induces an \emph{angular counting loss} of Bhabha events.

\subsubsection*{Initial State Radiation}

The Bhabha process is accompanied by emission of the initial-state radiation (ISR) that is nearly collinear with the initial particle momenta, as well as the final-state radiation (FSR) that is approximately collinear with the outgoing particle momenta. Since the ISR is nearly collinear with the beam axis, it misses the luminometer, so that the CM energy reconstructed from the detected particles is $E_{CM,rec} < E_{CM}$, and the corresponding spectrum can be represented as a generalized convolution of $\mathcal{B}(E_{CM})$ and the function $\mathcal{I}(x)$ describing the fractional CM energy loss due to the ISR,

\begin{equation}
\label{eq-isr-loss}
h(E_{CM,rec}) = \int\limits_0^{E_{\max}}{\mathcal{B}(E_{CM}) \frac{1}{E_{CM}} \mathcal{I}(\frac{E_{CM,rec}}{E_{CM}}) \ud E_{CM}}
\end{equation}

As the analysis of the luminosity spectrum in this work is limited to the energies above 80\% of the nominal CM energy, the function $\mathcal{I}(x)$ is taken to be approximately independent of $E_{CM}$.

In the frame of the two-electron system\footnote{Unless stated otherwise, electron always refers to electron or positron} after emission of ISR and before emission of the FSR, i.e. the \emph{collision frame}, the deflection angles in the collision are the same for both particles, according to the momentum-conservation principle. One can, therefore, define a unique \emph{scattering angle} $\theta^{coll}$.\footnote{In reality, ISR and FSR cannot be cleanly separated even theoretically, due to the quantum interference between them. Thus in practice, the collision frame is defined as the CM frame of the final electrons together with all radiation within a given tollerance angle with respect to the final electron momenta.} However, the loss of momentum due to ISR induces an additional relativistic boost in the outgoing particle angles in the same way as the Beamstrahlung. The detected angles are affected by the cumulative deformation from these two angular effects. 

Finally the outgoing particles are deflected under the influence of the EM field of the bunches, thus inducing a small additional angular counting loss termed Electromagnetic Deflection (EMD) effect. Synchrotron radiation may be emitted in this process as well.

\subsection{Event simulation methods}
\label{sec-sim}

To test the analysis procedures, Bhabha events in the bunch-collision were simulated with the modified Guinea-Pig software \cite{Sch96} using a method resembling that by C. Rimbault et al. in Ref. \cite{Rim07}: After generating the initial four-momenta of the colliding $e^- e^+$ pairs, the decision is made by Guinea-Pig whether the Bhabha scattering will be realized in the collision, based on the $1/s$ proportionality of the Bhabha cross section. If a Bhabha event is to be realized, the final four-momenta are picked from a file generated at the nominal CM energy $E_0$ by the BHLUMI generator \cite{Jad97}. The momenta are then scaled to the CM energy of the colliding pair $E_{CM}$, rotated to match the collision axis, and then boosted back to the lab frame. Finally the outgoing Bhabha electrons are tracked to simulate the electromagnetic deflection. 

The BHLUMI generator was run with the cuts on the polar angles in the lab frame $\theta_{min}^{lab} = 10 \text{\, mrad}$ and $\theta_{max}^{lab} = 200 \text{\, mrad}$. After generation, post-generator cuts were applied on the scattering angle $\theta^{coll}$ in the collision frame, and only events with $37 \text{\, mrad} < \theta^{coll} < 75 \text{\, mrad}$ were kept in the event file to be read by Guinea-Pig. The cross section corresponding to these selection criteria is 2.8 nb at 500 GeV, and 0.69 nb at 1 TeV.

The Guinea-Pig simulation was run with the standard parameter set from the ILC Technical Progress Report 2011 \cite{IDR11} as the basis for both the 500 GeV and the 1 TeV cases. Beside the standard parameter set, simulations were also performed with variations of individual beam-parameters, in order to determine the influence of the beam-parameter uncertainties on the performance of the presented methods. The simulated beam-parameter variations include:

\begin{itemize}
\item Symmetric bunch size variations by $\pm 10$ and $\pm 20\%$ and one-sided variations by +20\% in each of the three spatial directions
\item Symmetric charge variations by $\pm 10$ and $\pm 20\%$ and one-sided +20\% variation
\item Beam offset in x- and y-direction by up to one respective bunch RMS width. 
\end{itemize}

Earlier studies have shown that it is possible to control beam-parameter variations with a precision of 10\% or better \cite{Stahl05a, Grah08}. The scope of beam-parameter variations simulated here corresponds thus to a conservative expectation regarding the beam parameter stability. In total, 25 sets of beam parameters were simulated for each of the two energy options. In each simulation, one single beam parameter was varied with respect to the standard parameter set. Each of the simulations corresponds thus to a hypothetical situation when one beam parameter differs from its nominal value, and there are no variations of instantaneous luminosity within one simulation. In a real measurement, fluctuations in instantaneous luminosity can be expected. This fact implies that, in order to be applicable to the experimental situation involving undetected fluctuations of the beam parameters, the dependence of the final analysis results on the beam-parameter variation must be either negligible or linear to a good approximation. 

It should be noted that not all imaginable beam imperfections (such as banana shapes, rotations etc.) were covered in this analysis, and that physical correlations among the beam parameters were not taken into account. Reference values of beam parameters and their possible deviations and correlations are continuously evolving as the R\&D effort progresses, and in such a situation it seems unreasonable to attempt an exaustive analysis of their influence. The conservative range of magnitudes of the beam-parameter variations that were simulated serves thus as a safety margin against a too optimistic assessment of uncertainties.

%If this condition is satisfied, the analysis of any real measurement with change in instantaneous luminosity is equivalent to a linear combination of analysis results of several hypothetical measurements with parameters that are constant in time, but vary from measurement to measurement.

One bunch crossing was simulated in each case, and the Bhabha cross-section was multiplied by a factor in order to obtain sufficient statistic. The multiplication factor was fixed for all parameter sets within one energy case. Thus all the simulations within one energy case correspond to the same number of effective bunch crossings. As a consequence of the beam-parameter variation, the simulated effective luminosity varies between 0.5 and $1.4 \text{\, fb}^{-1}$ in the 500 GeV case, and between 1.2 and $5.6 \text{\, fb}^{-1}$ in the 1 TeV case. The number of Bhabha events in the simulations was between 1.5 and 4 million. 

The crab-crossing scheme was assumed, in which the beam-beam interaction can be seen as head-on collision in the system of the two colliding bunches. The effect of the lateral boost with respect to the lab frame is assumed to be properly handled by the positioning and the calibration of the detector. Thus the simulation was performed with head-on collision of the bunches, and the lateral boost was not considered in the analysis.

The interaction with the detector was approximated in the following way:
\begin{itemize}
\item The four-momenta of all particles within 5 mrad of the most energetic shower detected in the FV were summed together. The 5 mrad criterion corresponds closely to the Moli\`ere radius of the high-energy showers in the LumiCal \cite{Sad08}. The Beamstrahlung photons were not included as they are emitted close to the beam axis. For synchrotron radiation, the characteristic emission angles are of the order $1/\gamma$, which is smaller than $10^{-3}$ mrad for electron energies in the TeV range, therefore photon and electron four-vectors can be added.
\item The energy resolution of the LumiCal was included by adding random fluctuations to the final particle energies. The random fluctuations were sampled from the Gaussian distribution with energy-dependent standard deviation $\sigma_E/E = \sqrt{a^2/E + b^2}$. In existing analyses for both the ILC and CLIC variants of the LumiCal, the value of the stochastic term $a$ is found to be 0.21 \cite{Agu11, Schw11}, but there is no conclusive agreement on the value of the constant parameter $b$. In this work, two estimates $b = 0.011$ \cite{Agu11} and $b=0$ \cite{Schw11} were tested, and the effect of this choice on the final counting uncertainty was found to be negligible. The results and figures presented here correspond to the case $b = 0.011$.
\item The finite angular resolution of the LumiCal was included by adding random fluctuations to the final particle polar angles. Two values of $\sigma_\theta$ were tested. The first one is based on Ref. \cite{Abr07}, $\sigma_\theta = 1.5 \times 10^{-4}$, disregarding the energy dependence for simplicity. The second tested value is $\sigma_\theta = 2.2 \times 10^{-5}$, estimated in Ref. \cite{Sad08} for the advanced clustering method presented there. These values represent reasonable extremes for $\sigma_\theta$. The difference in the final counting uncertainty is very small or negligible, depending on the beam-beam correction method. The results and figures presented here correspond to the case $\sigma_\theta = 1.5 \times 10^{-4}$.
\end{itemize}

\begin{figure}
\centering
\includegraphics{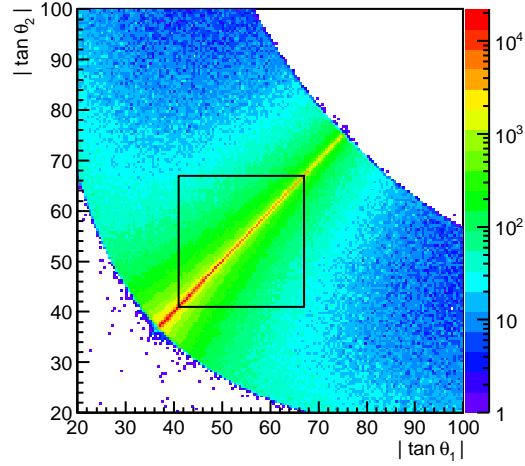}
\caption{\label{fig-2D}Scatter plot of the absolute values of polar angles of the outgoing Bhabha pairs in the Guinea-Pig simulation at 1 TeV. The thin black frame represents the angular limits of the FV. Events were generated with the scattering angle in the collision frame between 37 and 75 mrad. }
\end{figure}

Fig. \ref{fig-2D} shows the scatter plot of the absolute values of polar angles of the outgoing Bhabha pairs in the Guinea-Pig simulation at 1 TeV. The angular limits of the nominal fiducial volume (FV) of the LumiCal are represented by the thin black frame in the figure. In the ILD design of the LumiCal, the FV is defined by the limiting angles $\theta_{\min} = 41 \text{\ mrad}$ and $\theta_{\max} = 67 \text{\ mrad}$ \cite{ILD10, Abr10}. 

The central diagonal representing collinear events is immediately recognizable in the figure due to the high number of collinear events. The density of events falls of rapidly with increasing acollinearity.

\section{Correction methods}
\label{sec-corr}

In order to be able to properly define the counting bias in different analysis steps, and to separately treat different sources of bias, we adopt the following notation in this work:

The relative luminosity bias due to a given effect $\alpha$ is expressed as,

%under assumption that the CM frame of the colliding pair coincides with the lab frame.
% in absence of the beam-beam effects such that.

% In these methods, the relative counting biases $(\Delta N/N_{CM})_\alpha$ are determined by simulation of the beam-beam collision. The subscript $\alpha$ is provisionally used here to denote different effects inducing counting biases (see Sec. \ref{sec-phys}). The luminosity result is then corrected by the factor $1 - (\Delta N/N)_\alpha$ for each $\alpha$. The statistical uncertainty of the MC correction obtained here is denoted $(\sigma_N /N)$.

\begin{equation}
\label{eq-dnn}
\frac{\Delta L_\alpha}{L} = \frac{1}{L} \frac{N_{meas,\alpha} - N_{th,\alpha}}{\sigma_{Bh}} = \frac{\Delta N_\alpha}{N_{th}}
\end{equation}

Here the counting bias due to the effect $\alpha$ is $\Delta N_\alpha = N_{meas,\alpha} - N_{th,\alpha}$, where $N_{meas,\alpha}$ is the Bhabha count in presence of the effect $\alpha$, and $N_{th,\alpha}$ stands for the theoretically expected Bhabha count in the absence of the effect $\alpha$. Depending on the context, $N_{th,\alpha}$ may include bias from other beam-beam effects. This will be precised for each effect and each analysis method in its own section. The effective Bhabha cross section $\sigma_{Bh}$ is calculated with angular and energy cuts specified for each method in its own section, and $N_{th}$ represents the theoretical Bhabha count in absence of beam-beam effects. In the present analysis, $N_{th}$ is calculated by application of the angular and energy cuts defined in each section directly to the BHLUMI event sample.

The first of the correction methods presented here employs angular selection  optimized in such a way that the components of the beam-beam induced counting bias partially cancel out, and the resulting bias is nearly insensitive to the variations in the beam parameters. This method will be termed the \emph{compensation method}.

In the \emph{luminosity-spectrum method}, the counting loss is determined from the ratio of the tail to peak integrals in the experimentally reconstructed $\sqrt{s'}$ spectrum. 

In the \emph{collision-frame method}, the velocity of the collision frame of the Bhabha scattering is experimentally determined, and the corresponding loss of acceptance relative to the collision frame is calculated and corrected event by event \cite{Luk12}. The resulting count $N$ thus corresponds to the number of events with the scattering angle within the given limits in the collision frame, and the cross-section can be integrated within the same angular limits in the same reference frame.

The luminosity-spectrum method and the collision-frame method deal only with the component of the counting bias due to Beamstrahlung. The method to separately determine the EMD component of the counting bias will be presented at the end.

\subsection{The compensation method}
\label{sec-compensation}

\begin{figure}
\centering
\includegraphics{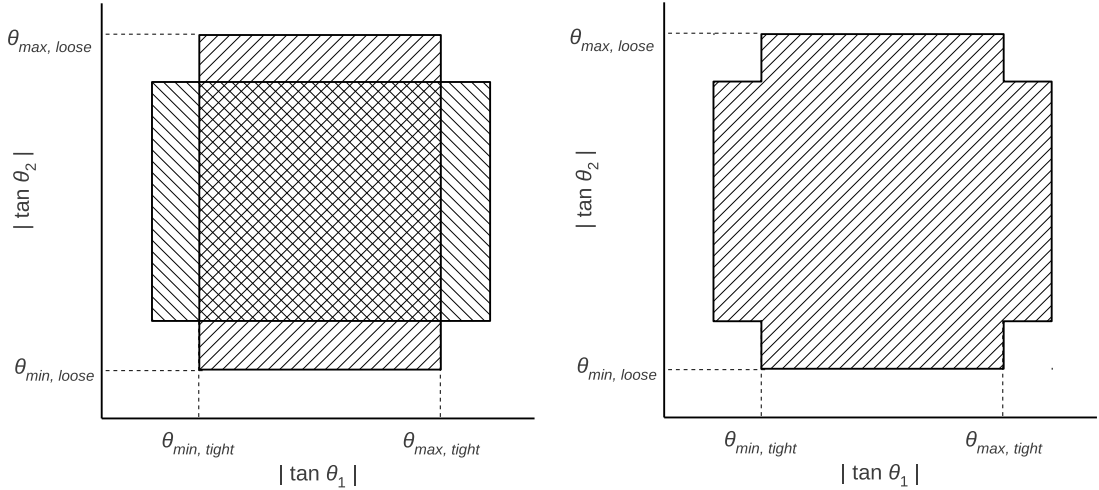}
\caption{\label{fig-asymm}Schematic representation of the angular selection algorithms used at LEP. Left: the OPAL and the ALEPH collaborations, right: the L3 collaboration}
\end{figure}

At LEP, various asymmetric selection cuts were employed \cite{Opal00, Aleph00, L3_00} to minimize the effect of the forward-backward asymmetry on the Bhabha counting rate, induced by the beam-beam effects, as well as by the longitudinal shifts of the beamspot and the alignment uncertainties. Similar techniques were proposed for ILC, as well \cite{Stahl05, Rim07}. In these techniques, two different angular cuts are applied at different side of the luminometer, one "tight" and one "loose". The tight cut typically corresponds to the FV of the luminometer, and the loose to the geometrical acceptance. Depending on which side of the Luminometer the tight cut is made, two asymmetric cuts on the polar angles, $\Xi_L(\theta^{lab}_1,\theta^{lab}_2)$ and $\Xi_R(\theta^{lab}_1,\theta^{lab}_2)$ are defined. In Refs. \cite{Opal00, Aleph00, Stahl05, Rim07}, the arithmetic average of the counts obtained with $\Xi_L$ and $\Xi_R$ is used (Fig. \ref{fig-asymm}, left). In that way, the effective relative weight assigned to the region corresponding to both particles being detected in the FV is twice as high as at the borders where only one particle is in the FV. In Ref. \cite{L3_00}, the selection criterion is defined as logical OR between $\Xi_L$ and $\Xi_R$, thus distributing the weight uniformly over a region shown in Fig. \ref{fig-asymm}, right. 

It has been shown that the application of the OPAL-type asymmetric selection algorithm effectively reduces the beam-beam effects at ILC as well \cite{Stahl05, Rim07}. This can be intuitively understood if one considers that all sources of forward-backward asymmetry displace the coordinates of the pair nearly in the direction perpendicular to the collinearity diagonal in the $|\tan{\theta_1}|$ vs. $|\tan{\theta_2}|$ space. The regions the most strongly affected by the counting loss due to this effect are in the vicinity of the lower-left and the upper-right corner in the diagram. Then the OPAL-type selection cuts allow for a partial recovery of the lost events in the border regions. 
%We may also note that the recovery of counts can be expected to be even more effective 
Similar results can be expected with the L3-type cuts (Fig. \ref{fig-asymm}, right). We also note that the upper-left and the lower-right corners do not play a significant role in the recovery of counts.

\begin{figure}
\centering
\includegraphics{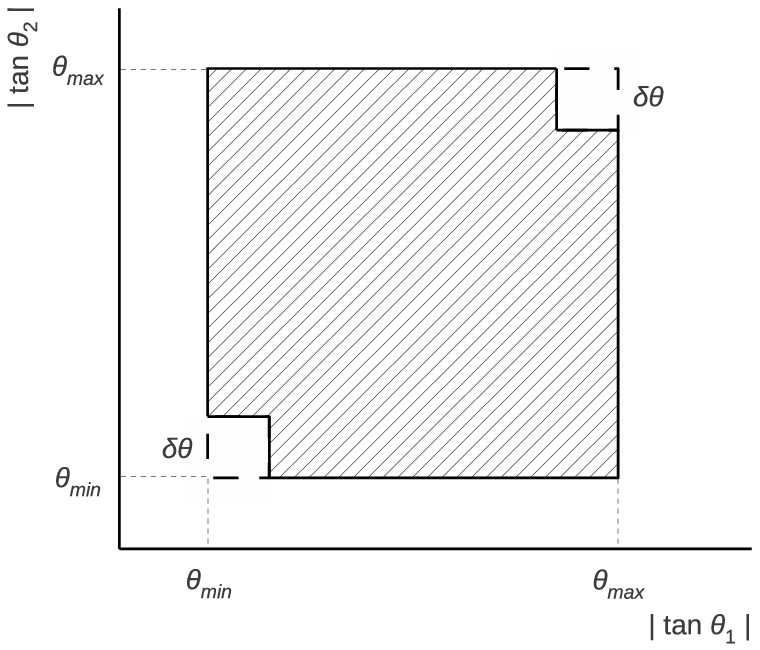}
\caption{\label{fig-dents}Schematic representation of the proposed angular selection algorithms. The solid line illustrates the modified selection criteria with the parameter $\delta \theta$}
\end{figure}

These considerations motivate the application of selection algorithms tailored specifically to optimize the counting loss. Such an algorithm can be constructed by requiring that both Bhabha particles be detected in the FV, except in the regions where both final polar angles $\theta^{lab}_1$ and $\theta^{lab}_2$ fall either in the interval $(\theta_{min}, \theta_{min} + \delta \theta)$ or in the interval $(\theta_{max} - \delta \theta, \theta_{max})$, as shown in Fig \ref{fig-dents}. The topology of the selection is similar to the selection employed at L3, but no events are accepted outside of the FV in this scheme. 

In addition to the angular cuts represented in Fig \ref{fig-dents}, the additional requirement that the sum of the detected particle energies be higher than 80\% of the nominal CM energy is applied in the present analysis.

\begin{figure}
\centering
\includegraphics{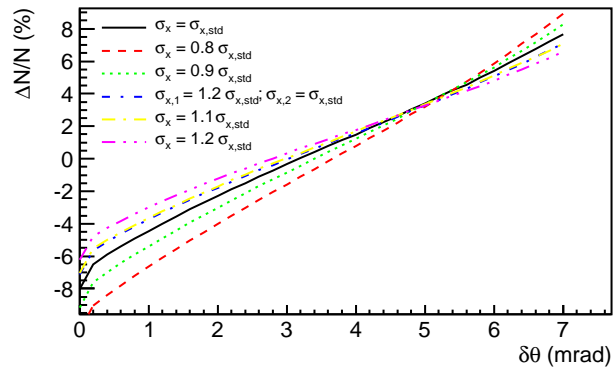}
\caption{\label{fig-optim}Total counting bias (Beamstrahlung and EMD) in the 1 TeV case as a function of $\delta \theta$ for five different $\sigma_x$ bunch widths ranging from 80 to 120 \% of the nominal $\sigma_x$.}
\end{figure}

The parameter $\delta \theta$ can be optimized using simulations. Fig. \ref{fig-optim} shows the total (Beamstrahlung and EMD) counting bias as a function of $\delta \theta$ for five different simulated bunch widths $\sigma_x$. The counting bias is in this case defined as $\Delta N = N_{meas}(\delta\theta) - N_{th}(\delta\theta)$, where $N_{meas}(\delta\theta)$ represents the measured number of events in the lab frame, within angular and energy cuts described above, and with all beam-beam effects included. The theoretical count in absence of beam-beam effects is represented by $N_{th}(\delta\theta) = \sigma_{Bh,\delta \theta} L$. In the present analysis, $N_{meas}(\delta\theta)$ was obtained by applying the described angular and energy cuts to the events produced in the beam-beam simulation described in Sec. \ref{sec-sim}, and $N_{th}(\delta\theta)$ was obtained by applying the same cuts directly to the BHLUMI event sample.

As seen in Fig. \ref{fig-optim}, for $\delta \theta = 0$, $\Delta L/L$ is negative. For given value of $\delta \theta$ depending on the beam parameters, the absolute value of $\Delta L/L$ is reduced to zero. However, the most attractive situation is the value of $\delta \theta$ for which the relative variation of the count due to the variation of the beam parameters is minimal. When the entire set of beam imperfections described in Sec. \ref{sec-sim} is taken into account (not shown in Fig. \ref{fig-optim}), taking $\delta \theta = 4.6 \text{ mrad}$ in the 500 GeV case minimizes the maximum counting variation with respect to the standard beam parameters to 2.6 permille. Similarly, in the 1 TeV case, $\delta \theta = 5.2 \text{ mrad}$ results in the maximum counting variation of 2.8 permille.

Despite the somewhat reduced geometric counting efficiency, the four-year integral luminosity at ILC is sufficient to ensure statistic uncertainty better than $1 \times 10^{-4}$ with this selection scheme.

\subsection{Luminosity-spectrum method}
\label{sec-spectrum}

\begin{figure}
\centering
\includegraphics{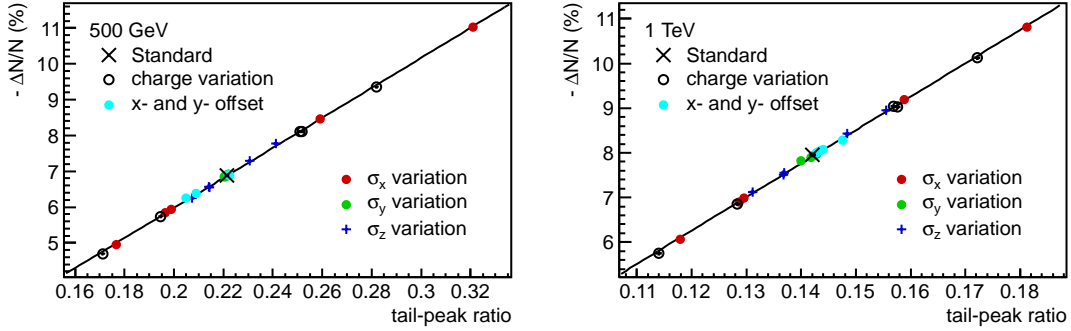}
\caption{\label{fig-tp-BS}Correlation of the tail-to-peak ratio in the energy spectrum reconstructed from acollinearity to the Beamstrahlung counting loss for a diverse range of beam imperfections based on the standard parameter set from Ref. \cite{IDR11}. Left - ILC at 500 GeV, right - ILC at 1 TeV}
\end{figure}

It has already been shown that the form of the $E_{CM}$ spectrum reconstructed from acollinearity of the outgoing pairs correlates with the counting loss due to Beamstrahlung \cite{Rim07}. To quantify the correlation, the ratio of the integral in the tail to the peak is used in this work as the parameter of the form of the spectrum. The correlation to the counting loss is determined by simulation. Fig. \ref{fig-tp-BS} shows the results at 500 GeV and 1 TeV for the whole range of simulated beam imperfections described in Sec. \ref{sec-sim}. 

The counting bias is defined as $\Delta N = N_{meas,BS} - N_{th}$, where $N_{meas, BS}$ is the number of events satisfying the angular- and energy selection cuts in the intermediate step in Guinea-Pig after Bhabha scattering, but before tracking in the EM field of the bunches (see Sec. \ref{sec-sim}). In this way, the Beamstrahlung counting loss is analyzed separately from EMD (see Sec. \ref{sec-EMD}). The angular selection includes the whole FV of the LumiCal on both sides, and the energy selection requires that the $E_{CM}$ reconstructed from acollinearity is higher than 80\% of the nominal CM energy.

In the 500 GeV case, the tail integral was calculated in the range from 400 to 480 GeV, and the peak integral from 480 GeV upwards. In the 1 TeV case, the tail was integrated from 800 to 940 GeV, and the peak from 940 GeV upwards. These energy ranges were obtained by optimization to ensure a correlation that is as linear as possible, when all types of beam imperfections are included. The linearity is important in order to minimize the uncertainty coming from the expected fluctuations of the bunch parameters in the real experiment. 

The correlation of the counting bias with the tail-to-peak ratio is characterized by the linear fit to the data presented in Fig. \ref{fig-tp-BS}. The fitted dependence can be used to correct the Beamstrahlung counting bias by measuring the tail-to-peak ratio of the reconstructed spectrum. A conservative estimate of the uncertainty of this correction is given by the worst deviation from the fitted function among the data points in Fig. \ref{fig-tp-BS}. This uncertainty is 0.7 permille of the total luminosity, in both the 500 GeV and the 1 TeV cases. The deviations for all points correspond well to the deviations expected from the statistical uncertainties of the simulated counting bias and the errors of the fit parameters. The robustness of the correlation of the tail-to-peak ratio to the Beamstrahlung component of the counting loss, regardless of the specific type of beam imperfection, gives this parameter credibility for accurate correction of the Beamstrahlung effect.

\subsection{Collision-frame method }
\label{sec-coll}

In the collision-frame method proposed in Ref. \cite{Luk12}, the analysis of the Bhabha count proceeds in three steps. In the first step, the influence of the Beamstrahlung on the effective angular acceptance is dealed with, in the second step, the ISR energy loss is deconvoluted from the energy spectrum of Bhabha events, and in the third step, small numerical correction is applied for the counting bias due to the LumiCal energy resolution. The separate correction of the EMD counting bias, which is of interest in the ILC case presented here, can be regarded as the fourth step. 

\subsubsection{Angular counting loss}

As already noted in Sec. \ref{sec-phys}, in the collision frame the deflection angles of both colliding particles are the same due to the momentum conservation principle. The movement of the collision frame with respect to the lab frame is responsible for the acollinearity leading to the angular counting loss. Since the kinematics of the detected showers correspond to the system after ISR, the velocity of the collision frame $\vec{\beta} _{coll}$ can be calculated to a good approximation from the measured polar angles. If $\beta_{coll}$ is taken to be collinear with the $z$-axis, the expressions for the boost of the Bhabha scattering angles into the lab frame give,

\begin{equation}
\label{eq-beta}
\beta_{coll} = \frac{\sin (\theta^{lab}_1 + \theta^{lab}_2)}{ \sin \theta^{lab}_1 + \sin \theta^{lab}_2 }
\end{equation}

Due to the longitudinal boost of the final particle angles, the effective acceptance of Bhabha events in the luminometer decreases with increasing $\beta_{coll}$. The effective limiting scattering angles $\theta^{coll}_{min}$ and $\theta^{coll}_{max}$ in the collision frame for a given $\beta_{coll}$ are obtained by boosting $\theta_{min}$ and $\theta_{max}$ into the collision frame. This allows calculating the event-by-event weighting factor to compensate for the loss of acceptance,

\begin{equation}
\label{eq-w}
w(\beta_{coll}) = \frac{\int\limits^{\theta_{max}}_{\theta_{min}} \frac{\ud\sigma}{\ud\theta} \ud\theta }{\int\limits^{\theta^{coll}_{max}}_{\theta^{coll}_{min}} \frac{\ud\sigma}{\ud\theta} \ud\theta}. 
\end{equation}

This part of the procedure corrects for the angular counting loss due to the Beamstrahlung and allows reconstructing the CM energy spectrum of the system after emission of the ISR (see Eq. \ref{eq-isr-loss}). The deconvolution of the ISR energy loss will be done in the next step.

To test the correction of the angular loss, histograms of $E_{CM,rec}$ reconstructed from the full kinematical information of the detected particles, were generated in the following way: 

\begin{description}
\item [Control histogram]: All events with the scattering angle in the collision frame $\theta^{coll}$ such that $\theta_{min} < \theta^{coll} < \theta_{max}$ are accepted. Therefore this histogram is not affected by counting losses due to Beamstrahlung and ISR. The energy spectrum is, however, affected by the ISR energy loss, and corresponds to the spectrum $h(E_{CM,rec})$ defined in Eq. \ref{eq-isr-loss}.
\item [Uncorrected histogram]: Events hitting the FV of the LumiCal in the lab frame.
\item [Corrected histogram]: Events hitting the FV of the LumiCal in the lab frame, stored with the weight $w$ calculated according to Eq. \ref{eq-w} 
\end{description}

To calculate the correction weight $w$, the approximate expression for the angular differential cross section $d\sigma / d\theta \approx \theta^{-3}$ was used. 

\begin{figure}
\centering
\includegraphics{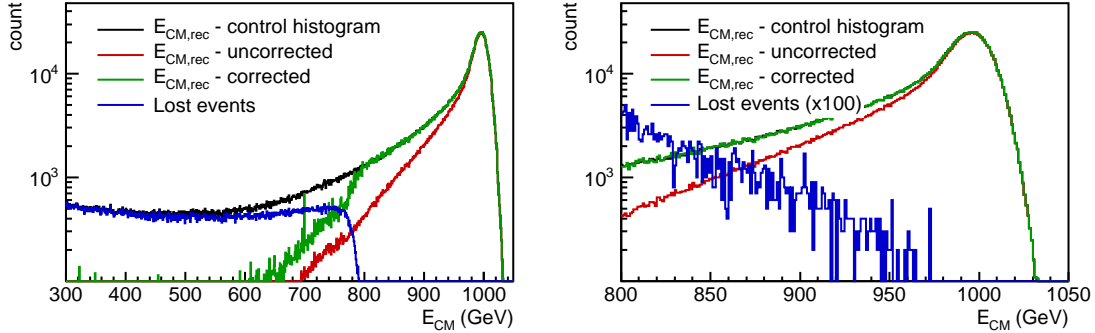}
\caption{\label{fig-BS-corr}Correction of the counting loss due to Beamstrahlung and ISR at 1 TeV. Left: whole spectrum; right: zoom on energies above 800 GeV. Black: Simulated control spectrum without counting loss due to Beamstrahlung and ISR; red: Reconstructed $E_{CM}$ spectrum affected by the counting loss; green: Reconstructed spectrum with correction for the counting loss due to Beamstrahlung and ISR; blue: events inaccessible to the correction}
\end{figure}

The results are shown in Fig. \ref{fig-BS-corr} for the 1 TeV case. The control spectrum is plotted in black, the spectrum affected by the counting loss in red, the corrected spectrum green. The blue line represents the events inaccessible to the correction due to the high value of $\beta_{coll.}$. The subsets of events characterized by $\beta_{coll.}$ above a certain treshold do not intersect the FV anymore \cite{Luk12}. These subsets cannot be corrected by this method. However, for such events, the Beamstrahlung-ISR energy loss is also above a certain minimum, so that they are only present in significant number below about 780 GeV. The presence of a small number of high-$\beta_{coll.}$ events at energies above 780 GeV is visible in the zoomed figure (Fig. \ref{fig-BS-corr}, right), where these events are scaled by a factor 100. In these events, $\vec{\beta} _{coll}$ has a relatively high radial component, due to the off-axis radiation before collision. For these events, the assumption that $\vec{\beta} _{coll}$ is collinear with the beam axis is not valid \cite{Luk12}. The mean relative bias due to such events to the peak integral above 80\% of the nominal CM energy is $(-1.424 \pm 0.007) \times 10^{-3}$ in the 1 TeV case, and $(-1.490 \pm 0.007) \times 10^{-3}$ in the 500 GeV case. 

The following is the list of sources of systematic uncertainty of the collision-frame method:

\begin{enumerate}
\item The assumption that  $\vec{\beta} _{coll}$ is collinear with the beam axis, 
\item The implicit assumption that the cluster around the most energetic shower always contains the Bhabha electron. In a small fraction of events, this is not the case, and the reconstructed polar angles $\theta_{1,2}^{lab}$ do not correspond to the final electron angles.
\item The use of the approximate angular differential cross section for the Bhabha scattering in the calculation of $w$,
\item Assumption that all ISR is lost, and all FSR is detected, in the calculation of $\beta_{coll}$ and $w$. 
\end{enumerate}

The first of these sources of uncertainty is responsible for the presence of the ``lost events'' in the peak. It induces a systematic bias that can be corrected by beam-beam simulation. The variation of this bias with the beam-parameter variation is smaller than 0.1 permille, indicating that the correction is reliable. The effect of the second uncertainty source is clearly seen when simulation with clustering around the electron is compared to the simulation with clustering around the most energetic shower. This effect produces a positive counting bias of $\lesssim 0.5$ permille both in the 1~TeV case and in the 500 GeV case. As the size of this effect depends on the details of the clustering and event-reconstruction procedure, no correction will be performed for it at this stage. The remaining two effects are not significant compared to the statistical uncertainty of the present analysis.

After correcting for the lost events, the average counting bias after correction was calculated for both energy options, for the entire set of simulations with variations of beam parameters (Sec. \ref{sec-sim}), taking the inverse square statistical uncertainty of the deviation as the weight. The average bias at 500 GeV is $(+0.39 \pm 0.09) \times 10^{-3}$, while at 1 TeV the average bias is $(+0.67 \pm 0.10) \times 10^{-3}$.

\subsubsection{ISR energy loss}

The histogram obtained in the previous step is an accurate model-independent reconstruction of the CM energy spectrum $h(E_{CM,rec})$ of Bhabha events with ISR energy loss, defined in Eq. \ref{eq-isr-loss}. To obtain the Bhabha CM energy distribution $\mathcal{B}(E_{CM})$, the ISR energy loss should be deconvoluted from $h(E_{CM,rec})$. This deconvolution can be performed using the theoretical form of the distribution $\mathcal{I}(x)$ of the ISR fractional energy loss, and by solving the system of linear equations resulting from the discretization of Eq. \ref{eq-isr-loss}, as described in \cite{Luk12}. To obtain the function $\mathcal{I}(x)$, the distribution of $x = E_{CM,rec}/E_{CM}$ was taken from the BHLUMI file, and the beta distribution was fitted to it for $x > 0.8$.

\begin{figure}
\centering
\includegraphics{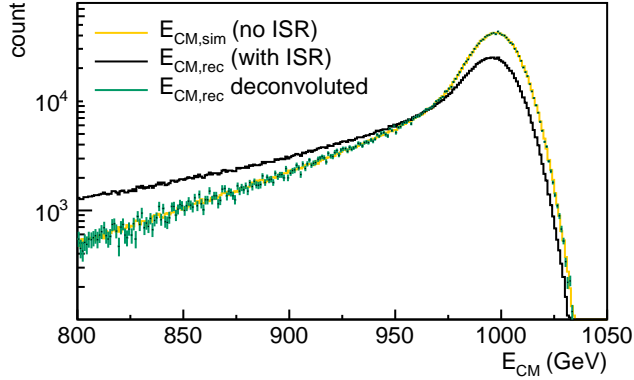}
\caption{\label{fig-deconv}Deconvolution of the ISR deformation of the luminosity spectrum. Yellow: the control histogram -- simulated $E_{CM}$ before emission of ISR, smeared with a normalized Gaussian; black: the histogram affected by the ISR energy loss -- reconstructed $E_{CM}$ from the detected showers, green: deconvoluted spectrum.}
\end{figure}

The results of the deconvolution are shown in Fig. \ref{fig-deconv}. In a manner similar to the previous section, three histograms were filled: 

\begin{description}
\item[Control histogram] (yellow line) contains simulated CM energies before ISR emission. At filling, this histogram was smeared by adding small Gaussian random fluctuations to the energy of the event, in order to match the effect of the energy resolution of the LumiCal on the deconvoluted histogram. In this way, the uncertainty of the deconvolution is treated separately from the counting uncertainty due to the energy resolution.
\item[Histogram with ISR energy loss] (black line) contains CM energies $E_{CM,rec}$ reconstructed from the final-state kinematics, with inclusion of the LumiCal energy resolution. 
\item[Deconvoluted histogram] (green points with error bars) was obtained by deconvolution of the ISR energy loss \cite{Luk12}.
\end{description}

The deconvoluted histogram obtained here corresponds to the CM energy spectrum of Bhabha events $\mathcal{B}(E_{CM})$. The luminosity spectrum can now be obtained in the top 20\% of the CM energy by normalizing $\mathcal{B}(E_{CM})$ bin-by-bin with the Bhabha cross section $\sigma(E_{CM})$ calculated for each CM energy bin. The calculation of the cross section should proceed by integration in the angular range $\theta_{min} < \theta^{coll} < \theta_{max}$ in the collision frame, and in the entire range of fractional ISR energy loss $ 0 < x < 1$. 

The uncertainty estimate of the deconvolution procedure alone for the integral luminosity in the upper 20\% of the spectrum is given by the relative integral difference between the deconvoluted and the control spectrum in the upper 20\%. This uncertainty is $(+0.81 \pm 0.22) \times 10^{-3}$ at 1 TeV, and $(+0.35 \pm 0.21) \times 10^{-3}$ at 500 GeV.

Not included in this uncertainty estimate is the effect of the simplifying assumption that the form of $\mathcal{I}(x)$ is independent of $E_{CM}$. This assumption was used to generate the BHLUMI event file for the simulation, as well as to extract the shape of $\mathcal{I}(x)$ for the deconvolution. Thus its effect cancels out in the comparison of the deconvoluted and the control histogram.

\subsubsection{Energy resolution}

Since the full energy information is used to determine the luminosity spectrum, the energy resolution of LumiCal induces a bias in the Bhabha count by asymmetric redistribution of events around the CM energy cut because of the slope in the form of the spectrum at the cut energy. This can be corrected by integration of the fitted parametrized form of $\mathcal{B}(E_{CM})$. The correction is obtained as the relative difference of the function integral when the fitted resolution parameter is replaced by zero. When the cut is made at 80\% of the nominal energy, the size of this correction is between 1 and $4 \times 10^{-4}$. It has been shown in Ref. \cite{Luk12} that the energy-resolution effect can be corrected to better than $1 \times 10^{-4}$.

%The result of the luminosity measurement with this method is the number of events $N_{coll.}$ with scattering angles between $\theta_{\min}$ and $\theta_{\max}$ in the collision frame. The same limits can be used for the cross-section integration in the same reference frame. The initial CM energy supplied to the generator for the cross-section calculation should be sampled from the realistic form of the intrinsic luminosity spectrum $\mathcal{L}(E_{CM})$ (See Sec. \ref{sec-phys}). The form of $\mathcal{L}(E_{CM})$ can be determined either by simulation or by deconvolution of the ISR energy-loss from the measured spectrum. 

\subsection{Angular losses due to the EMD}
\label{sec-EMD}

The EMD shifts the polar angles of the outgoing particles consistently towards smaller angles. Since the Bhabha cross section is monotonously decreasing with the polar angle, the net effect of the EMD is a decrease in the Bhabha count. The relative counting bias is defined as $(N_{meas,EMD} - N_{th,EMD})/N_{th}$, where $N_{meas,EMD}$ stands for the measured number of events when all beam-beam effects, including EMD, are taken into account, and $N_{th,EMD}$ is the simulated number of events when all effect \emph{except the EMD} are taken into account. 
%$N_{th}$ stands for the theoreticaly expected number of events in absence of any beam-beam effects.

The EMD bias is equivalent to the effect of a parallel shift in the angular limits $\theta_{min}$ and $\theta_{max}$ of the FV in the opposite direction by an \emph{effective mean deflection} angle $\Delta \theta$. 

\begin{equation}
\label{eq-emd}
\frac{\Delta L_{EMD}}{L} = \frac{1}{N} \frac{dN}{d\theta} \Delta \theta
\end{equation}

The value of $\Delta \theta$ can be obtained by dividing the relative EMD counting loss obtained in the beam-beam simulation by the simulated or measured quantity $x_{EMD} = \frac{1}{N} \frac{dN}{d\theta}$, where $N$ is the Bhabha count in the FV, and $d\theta$ denotes a parallel infinitesimal shift of both $\theta_{min}$ and $\theta_{max}$. 

\begin{figure}%[hb]
\centering
\includegraphics[width=85mm]{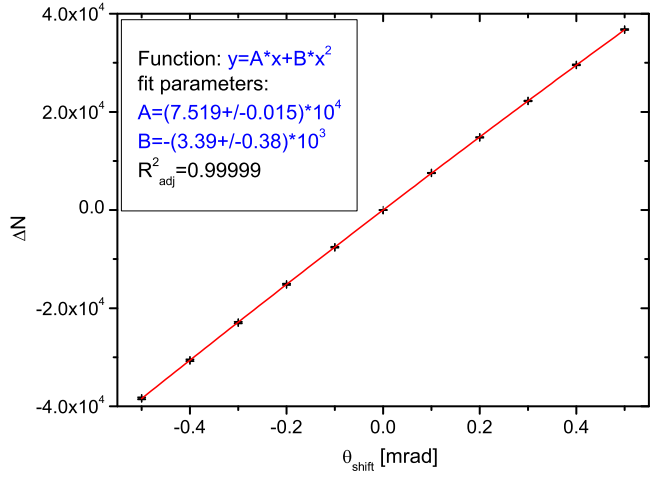}
\caption{\label{fig-dNdth}Fit of the $(dN/d\theta)_{sim}$ for the standard set of beam parameters at 1 TeV.}
\end{figure}

Fig. \ref{fig-dNdth} shows the fit of the $(dN/d\theta)_{sim}$ for the standard set of beam parameters at 1 TeV. $\Delta N = N_{shift} - N_{FV}$ is the difference in counts in the shifted angular cut $(\theta_{min} + \theta_{shift}, \theta_{max} + \theta_{shift})$, and the FV $(\theta_{min}, \theta_{max})$. The statistical errors of $\Delta N$ were estimated as $\delta(\Delta N) = \sqrt{n_{shift} + n_{FV}}$, where $N_{shift} = N' + n_{shift}$, $N_{FV} = N' + n_{FV}$ and $N'$ is the number of events inside the intersection of the FV with the shifted angular cut. The slope at zero is $dN/d\theta = (7.519 \pm 0.015)\times 10^{4} \text{\, mrad}^{-1}$. The number of counts in the FV in the simulation was $1.37 \times 10^{6}$, and the counting bias was $\Delta L_{EMD}/L = (-1.068 \pm 0.028) 10^{-3}$. The resulting effective mean deflection is $\Delta \theta_\text{1TeV} = (0.0194 \pm 0.0006) \text{\, mrad}$. 

In the 500 GeV case, the same procedure gives $\Delta \theta_\text{500GeV} = (0.0426 \pm 0.0009) \text{\, mrad}$. 

\begin{figure}
\centering
\includegraphics[width=70mm]{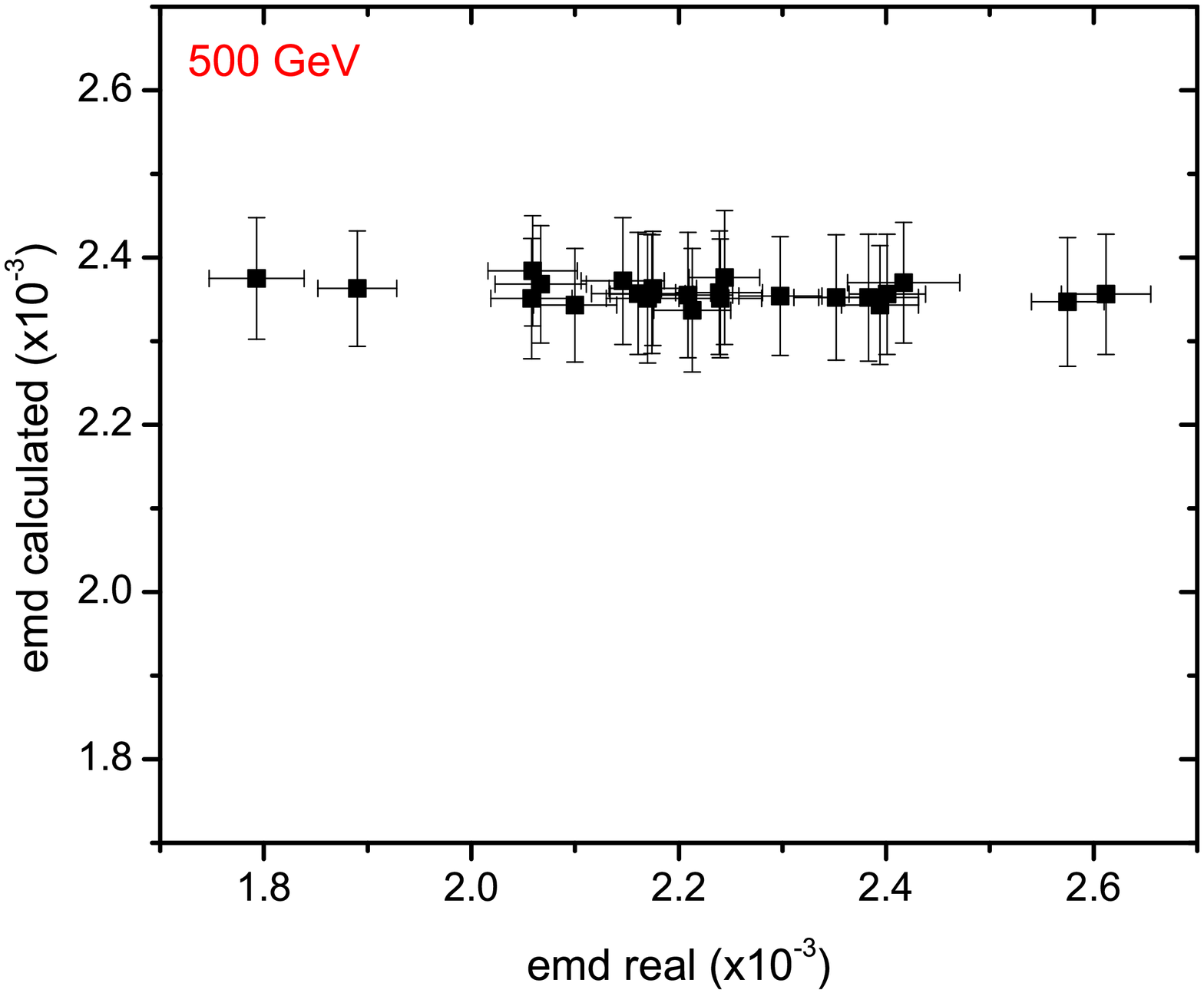}
\includegraphics[width=70mm]{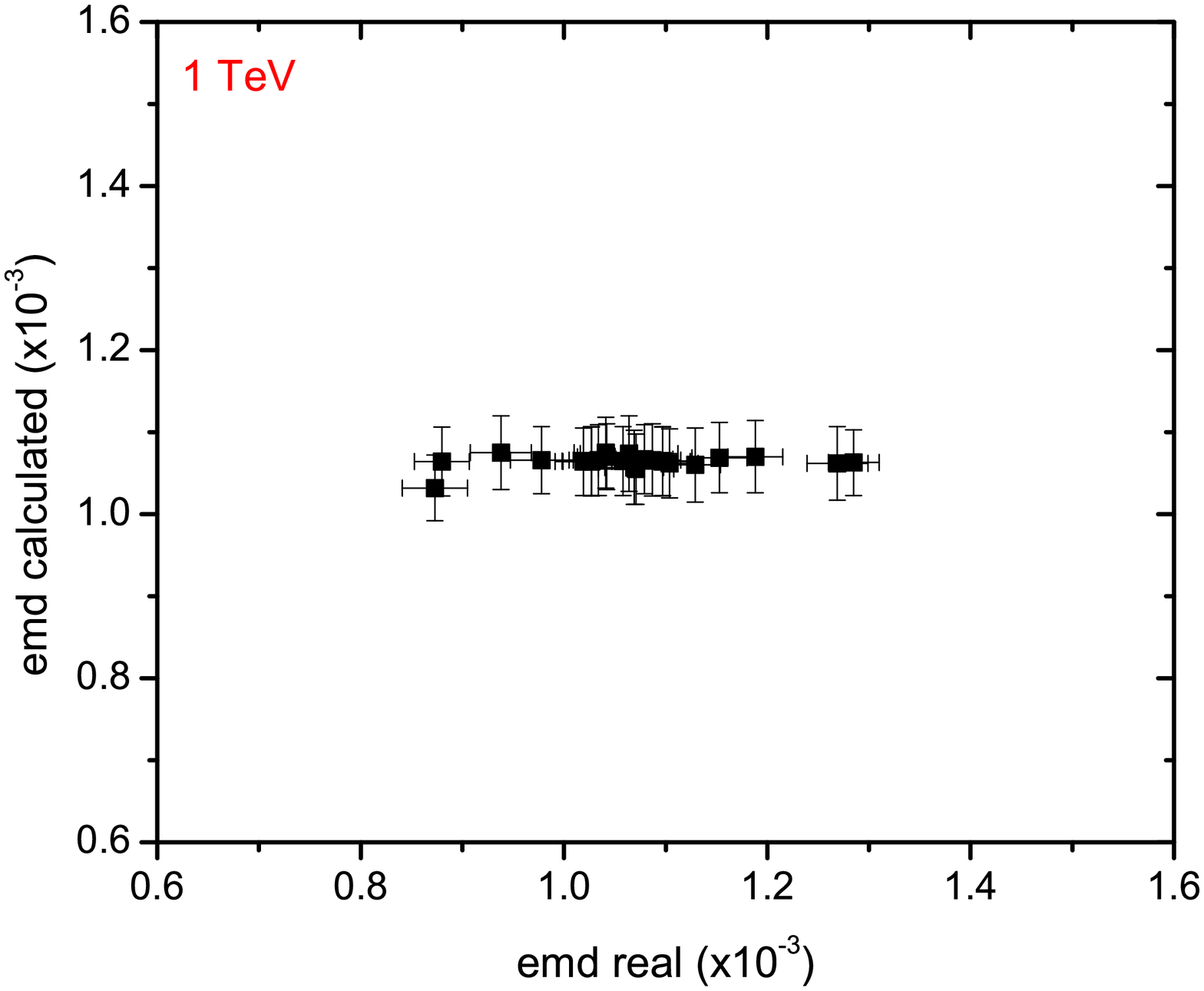}
\caption{\label{fig-emd}Scatter plot of the calculated EMD counting loss from $(dN/d\theta)_{sim}$ with a fixed $\Delta \theta$ value derived for the nominal beam parameters, against the "real" EMD obtained directly from the difference in counts for a diverse range of beam imperfections, including the bunch geometry, as well as the bunch charge variations. Left: the 500 GeV case, right: the 1 TeV case}
\end{figure}

If the simulated value of $\Delta \theta$ is returned to Eq. \ref{eq-emd}, it can be used in the experiment to correct for the EMD counting loss. The precision of this correction depends on the precision with which the beam parameters are known. Fixed value of $\Delta \theta$ obtained in the simulation with the nominal beam parameters is always used. Fig \ref{fig-emd} shows the scatter plot of the values of the EMD counting loss obtained using the fixed $\Delta \theta$ versus the "real" EMD loss $(N_{meas,EMD} - N_{th,EMD})/N_{th}$ for each set of beam parameters described in Sec. \ref{sec-sim}. No correlation between the calculated EMD and the beam parameters is seen that could be used to further refine the result. Thus at present only one nominal value of the EMD counting bias can be used for the correction.

In the worst case the error of this estimate is $\pm 5\times 10^{-4}$ of the total luminosity at 500 GeV, and $\pm 2\times 10^{-4}$ at 1 TeV. We take these worst-case uncertainties as a conservative estimate of the uncertainty of the EMD correction. This uncertainty is a factor 5 smaller than the uncorrected EMD counting bias itself. If the beam parameters are known with better precision than 20\% (see Ref. \cite{Grah08}), the residual uncertainty will be correspondingly smaller.

The results presented here refer to the selection criteria applied in the luminosity-spectrum method. With the collision-frame method, the size of the uncorrected EMD effect is slightly different, but the uncertainties after correction are the same.

\section{The collision-frame method and additional selection criteria}
\label{sec-add-cuts}

Because of the presence of background that can be falsely identified as Bhabha events in the luminosity measurement at a linear collider, there is a strong interest to apply additional cuts to limit the measurement to the region of phase space that is as free from background as possible. Several types of cuts are already well known from previous experiments. Beside the mentioned asymmetric polar-angle cuts, cuts on acoplanarity, acollinearity, minimum and average energy have been shown to effectively remove background \cite{Opal00}. Recently, the energy balance cut was also proposed \cite{Mila11}. 

In Sec. \ref{sec-corr}, it has been shown that the collision-frame method offers the possibility to reconstruct $\mathcal{L}(E_{CM}$ above 80\% of the nominal energy in independently of the knowledge of the beam parameters. However, when additional selection cuts are considered, the independence of the beam parameters can be preserved only if the selection cuts are either based on the Lorentz-invariant quantities, or on quantities that are defined in the same reference frame in the experiment as well as in the cross-section integration. These conditions are satisfied in practice to a good approximation in two cases presented in the following.

\subsection{Acoplanarity} 
\label{sec-dphi}

The acoplanarity criterion is defined as $|\pi - |\phi_1 - \phi_2|| < \Delta\phi_{max}$. Although this criterion is based on the azimuthal angles of the detected particles, which are in general not Lorentz invariant, this criterion is to an excellent approximation invariant with respect to the boost between the CM frame of the Bhabha event and the laboratory frame. The Beamstrahlung is so strongly forward-peaked that the corresponding boost nearly coincides with the beam axis. Beside this, the acoplanarity criterion suppresses events that have radiated significant off-axis ISR before collision. Therefore, it can be expected that the acoplanarity criterion reduces the fraction of lost events (Sec. \ref{sec-coll}).

In order to quantify these arguments, the collision-frame method was tested for the standard ILC parameter set with application of the acoplanarity criterion with several different values of $\Delta\phi_{max}$. The results are shown in Tab. \ref{tab-dphi}. As can be seen from the table, the final numerical precision as well as the statistics are not affected by the application of the acoplanarity criterion. However, the fracton of lost events is reduced several times when the acoplanarity limit is set to 5 or 10 degrees.

It can be concluded that the acoplanarity criterion can be optimized for the needs of background reduction. The side effect of it is that it also reduces the fraction of lost events, thus enhancing the overall reliability of the measurement method.

\begin{table}
\caption{\label{tab-dphi}Relative counting uncertainty after correcting the Beamstrahlung angular counting loss and the lost events, relative magnitude of correction for the lost events, and the total statistic in the collision-frame method with application of the acoplanarity criterion for different values of $\Delta\phi_{max}$.}
\begin{center}
\begin{tabular}[c]{@{}    l         r @{$\pm$} l  r@{$\pm$} l  c }
$\Delta\phi_{max} (^\circ)$         & \multicolumn{2}{c}{$\frac{\Delta N_{BS}}{N}$} & \multicolumn{2}{c}{$\frac{\Delta N_{lost}}{N}$} & $N$   \\
                                  & \multicolumn{2}{c}{$10^{-3}$} & \multicolumn{2}{c}{$10^{-3}$} & $10^{6}$ \\
\hline
180 (any acoplanarity accepted) &  +0.98&0.48 & 1.44&0.03 & 1.734  \\
57.3                              &  +0.76&0.48 & 1.23&0.03 & 1.723  \\
10                                &  +0.73&0.49 & 0.61&0.02 & 1.674  \\ 5                                 &  +0.71&0.47 & 0.36&0.02 & 1.637  \\ 
\hline
\end{tabular}
\end{center}
\end{table}

\subsection{Reduction of the acollinearity and the average energy ranges by the cut on CM energy}
The average energy and the acollinearity criteria are not Lorentz invariant, and are based on quantities that are severely affected by the longitudinal boost. However, these criteria can be effectively replaced by the cut on the reconstructed CM energy, which is strongly correlated both with the average energy of the two detected particles, and anticorrelated with the acollinearity. As already shown in Sec. \ref{sec-coll}, events with high reconstructed CM energy and high acollinearity are rare, and correspond to events with high off-axis radiation before collision. 

\begin{figure}
\centering
\includegraphics{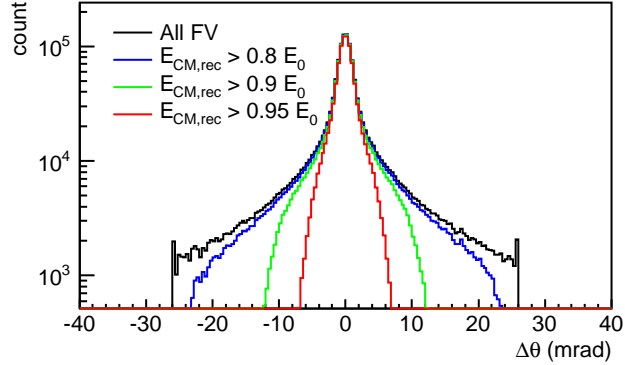}
\caption{\label{fig-dth}Distribution of $\Delta \theta = \theta_2 - \theta_1 - \pi$ for several different cuts on $E_{CM}$. Black line: no $E_{CM}$ cut -- all events in FV accepted; blue: events with $E_{CM} > 0.8 E_0$; green: events with $E_{CM} > 0.9 E_0$; red: $E_{CM} > 0.95 E_0$. Event weighting from the collision-frame method was applied.}
\end{figure}

As an illustration of the correspondence of the CM energy cut with the acollinearity cut, Fig. \ref{fig-dth} shows the distribution of the parameter $\Delta \theta = \theta_2 - \theta_1 - \pi$ for several different cuts on $E_{CM}$. Event weighting from the collision-frame method was applied. Without the $E_{CM}$ cut, $\Delta \theta$ is limited only by the requirement that the event falls within the FV. With the $E_{CM}$ cut, the range of $\Delta \theta$ is progressively smaller when the $E_{CM}$ cut is made nearer the peak, although the edges of the range are not sharp, as they would have been if the cut was made on $\Delta \theta$. In the tests in Sec. \ref{sec-coll} and \ref{sec-dphi}, the $E_{CM}$ cut was always made at 80 \% of the nominal energy. The exact position of the cut can be optimized if required for the background suppression.

\section{Conclusions}

Three methods for dealing with the beam-beam effects in the luminosity measurement were presented and tested using Monte Carlo event simulations. In the compensation method, the angular selection cuts are defined in such a way that the total counting bias (Beamstrahlung + EMD) is almost insensitive to the beam-parameter variations, and can be directly corrected by MC simulation. In the remaining two methods, the EMD component of the bias is determined separately. The counting bias without correction, as well as the estimated precision of the correction for the three presented methods is listed in Tab. \ref{tab-unc}. The listed uncertainties represent the final uncertainties after correcting for both the Beamstrahlung and the EMD effects in all cases. In the compensation method, both effects are treated at the same time, resulting in a single uncertainty value. In the case of the luminosity-spectrum methods, the final uncertainty is calculated as the quadratic sum of the uncertainties of the Beamstrahlung and the EMD corrections. The collision-frame method contains four analysis steps, each introducing an uncertainty of its own. The first step consists of the correction of the angular counting losses based on $\beta_{coll}$ and the MC correction of the lost events. The second step is the ISR deconvolution, the third step is the correction of the counting bias due to the energy resolution, and the fourth step is the correction of the EMD bias. Simulation results indicate that the biases after correction in the first two steps are systematically positive. Therefore, the uncertainties of the first two steps are linearly summed together. After that the uncertainties of the third and the fourth steps are added quadratically.
%The values of the final uncertainties are predominantly induced by the uncertainties of the beam parameters listed in Sec. \ref{sec-sim}.

\begin{table}[ht]
\caption{\label{tab-unc}Relative counting uncertainties after correction of the beam-beam effects using different methods.}
\begin{center}
\begin{tabular}[c]{@{}       p{8cm}                c             c   }
Method                                      & 500 GeV     & 1 TeV       \\
                                            & ($10^{-3}$) & ($10^{-3}$) \\
\hline\hline
Uncorrected bias                            &     -128    & -140    \\
\hline
Compensation                                &      2.6    &  2.8     \\ \hline
Luminosity spectrum and EMD correction      &      0.8    &  0.6  \\ \hline
Collision-frame method with corrections 
\newline for the lost events and EMD (add $\sigma_{Eres}$)     &      0.9  & 1.5     \\ \hline
\end{tabular}
\end{center}
\end{table}

The presented methods allow reaching precision of the order of permille or better in the luminosity measurement at ILC for both 500 GeV and 1 TeV, in the experimental situation where the key bunch parameters are known with a precision of 20\%. It may be noted that not all imaginable beam imperfections were covered by this analysis, and that physical (anti-) correlations among the beam parameters were not taken into account. On the other hand, the magnitude of the beam-parameter variations that were simulated is at least a factor 2 worse than the beam-parameter uncertainties expected from dedicated studies \cite{Stahl05a, Grah08}. Thus the final uncertainties presented here can be regarded as conservative.

The proposed EMD correction has the weakness that the effective deflection $\Delta \theta$ is not directly measured, but has to be estimated from simulation. For this, knowledge of beam parameters is necessary, and the method is sensitive to their uncertainties. However, as the EMD component of the bias is small at the outset, the final uncertainty is acceptable.

In the case of the collision-frame method, the choice of additional selection cuts for background removal is limited to cuts expressed in quantities that are invariant under the Lorentz boost from the CM frame to the lab frame. Two examples of such cuts were tested. In particular, the acoplanarity cut improves the condition related to the high-energy "lost events" due to the off-axis component of the ISR.

The collision-frame method results in the number of events $N$ in the angular range $(\theta_{min}, \theta_{max})$ defined in the collision frame. The same limits can be used for the cross-section integration in the same reference frame. Thus the luminosity expression (Eq. \ref{eq-luminosity}) can be written in a way that is essentially insensitive to the variation of the CM frame induced by the beam-beam effects. In this way, beam-beam simulation remains necessary only for the correction of the EMD effect, and of the high-energy "lost events". Both these corrections are of the order of 1 or 2 permille, while the main component of the beam-beam related counting bias is resolved in a robust way, independent of the beam-beam simulation and of the knowledge of the beam parameters. In addition, the luminosity spectrum in the range above 80\% of the nominal energy is accurately reconstructed. As a downside, integral luminosity only above ca 80\% can be accurately measured using this method. For studies of processes with threshold far below the nominal CM energy, results from other techniques, such as the luminosity-spectrum method may be more relevant.

\begin{table}[ht]
\caption{\label{tab-all-unc}Systematic uncertainties in luminosity measurement.}
\begin{center}
\begin{tabular}[c]{@{}       p{8cm}                c             c   }
Source of uncertainty                       & 500 GeV     & 1 TeV       \\
                                            & ($10^{-3}$) & ($10^{-3}$) \\
\hline\hline
Bhabha cross section \cite{Arb96}       &     0.6    & 0.6    \\
\hline
Beam-beam effects 
\newline(collision-frame method)       &      0.9    &  1.5     \\ \hline
Polar-angle resolution and bias \cite{Abr10} &     0.16   &  0.16  \\ \hline
Energy resolution \cite{Abr10}              &      0.1    &  0.1  \\ \hline
Energy scale \cite{Abr10}                   &      1      &    1  \\ \hline
Beam polarization  \cite{Abr10}             &      0.19   &  0.19    \\ \hline
Physics background \cite{Mila13}            &      2.2    &  0.8    \\ 
\hline\hline
Total                                       &      2.6    &  2.1    \\
\end{tabular}
\end{center}
\end{table}

Tab. \ref{tab-all-unc} lists the remaining sources of uncertainty in luminosity measurement. For the beam-beam correction, the uncertainty of the collision-frame method is given. The uncertainties due to the polar-angle resolution and the energy resolution are listed as well, although they are contained in the uncertainty of the beam-beam correction presented here. Their separate addition or not does not noticeably change the total uncertainty.

In the collision-frame method, the Bhabha cross section for the normalization of the measurement is supposed to be calculated without cuts on the final polar angles in the lab frame. The angular selection is made on the scattering angle. Since, in addition, the ISR deconvolution is performed on the experimental spectrum, the cross-section integration should proceed over the entire range of the outgoing particle energies. This, so far simplifies the radiative corrections required in the cross-section calculation. On the other hand, integration should be performed over the measured range of the initial CM energies, using the measured luminosity spectrum. This does not represent a particular difficulty. However, the deconvolution of the ISR energy loss in the luminosity spectrum now contains the theoretical uncertainty of the distribution $\mathcal{I}(x)$. Thus, overall, the same sources of theoretical uncertainty are present in the luminosity measurement as in previous studies \cite{Arb96, Jad03}, but they influence the result in a different way. Determination of the final theoretical uncertainty would require a study on its own. Since the required Monte-Carlo techniques cover the similar angular range as in the LEP calculations, we assume that the uncertainty will not be larger than at LEP \cite{Arb96}, although arguments have already been put forward that even better precision will be possible in the future \cite{Jad03}.

% The accolinearity of the final particles is, however, limited by the requirement that the CM energy of the event is higher than 80\% of $E_0$. Thus the integration of the cross section is performed in the region of the phase space in which the accuracy of the Bhabha generators has already been thoroughly studied. It is therefore safe to assume that the cross-section uncertainty is not higher than the values found in the literature.

The energy-scale and the beam-polarizaton uncertainties were discussed in Ref. \cite{Abr10}. The physics background was simulated in Ref. \cite{Mila13}. The theoretical uncertainty of the background cross section is unknown, due to the matrix-element singularities for the four-fermion processes with fermions escaping at small angles. Therefore, the full-size effect is cited for the physics background. One may note however, as stated in \cite{Abr10} based on results from LEP \cite{Poz07}, that the physics background could be corrected with an uncertainty of 40\% of its size.

The final uncertainty is 2.6, respectively 2.1 permille in the 500 GeV and 1 TeV cases. This satisfies the requirement for the largest part of the Physics programme at the ILC. However, for high-precision measurements such as the Giga-Z programme, precision of $10^{-4}$ is required \cite{RDR07}. Uncertainties presented here are, however, not final, and may be refined towards this goal as more precise knowledge becomes available on beam-parameter physical correlations, the cross section of the physics background, as well as with further refinement of the correction methods.

%Using methods described in this note, the beam-beam related angular losses are well controlled, and  are not a critical source of uncertainty. 

\bibliographystyle{ams}
\bibliography{lumi}

\providecommand{\noopsort}[1]{}\providecommand{\singleletter}[1]{#1}%
\begin{thebibliography}{10}

\bibitem{Luk12}
S. Luki\'c
\newblock {\it {Correction of systematic uncertainties due to beam-beam effects
  in luminosity measurement in LumiCal at CLIC}}
\newblock Technical Report {LCD-Note-2012-008}, {Linear Collider Detector
  project, CERN}, 2012 (arXiv:1301.1449 [physics.acc-ph]).

\bibitem{Jad03}
S. Jadach
\newblock {\it Theoretical error of luminosity cross section at {LEP}}, 2003
  (arXiv:hep-ph/0306083).

\bibitem{Yok91}
K. Yokoya and P. Chen,
\newblock {\it Beam-beam phenomena in linear colliders},
\newblock in Frontiers of Particle Beams: Intensity Limitations, Vol. 400 of
  {\it Lecture Notes in Physics} Springer Verlag, 1991, pp.  415--445
\newblock ({US-CERN} School on Particle Accelerators).

\bibitem{Sch96}
D. Schulte,
\newblock {\it Study of Electromagnetic and Hadronic Background in the
  Interaction Region of the {TESLA} Collider},
\newblock PhD thesis, Hamburg University, 1996.

\bibitem{Opal00}
{OPAL Collaboration},
\newblock {\it {Precision luminosity for Z0 line shape measurements with a
  silicon tungsten calorimeter}},
\newblock Eur.Phys.J.
\newblock {\bf  C14} (2000), 373--425.

\bibitem{Aleph00}
{ALEPH Collaboration},
\newblock {\it {Measurement of the Z resonance parameters at LEP}},
\newblock Eur.Phys.J.
\newblock {\bf  C14} (2000), 1--50.

\bibitem{L3_00}
{L3 Collaboration},
\newblock {\it {Measurements of cross-sections and forward backward asymmetries
  at the $Z$ resonance and determination of electroweak parameters}},
\newblock Eur.Phys.J.
\newblock {\bf  C16} (2000), 1--40.

\bibitem{Stahl05}
A. Stahl
\newblock {\it {Luminosity measurement via {Bhabha} scattering: Precision
  requirements for the luminosity calorimeter}}
\newblock Technical Report LC-DET-2005-004, SLAC, 2005.

\bibitem{Rim07}
C. Rimbault, P. Bambade, K. M\"{o}nig, and D. Schulte,
\newblock {\it Impact of beam-beam effects on precision luminosity measurements
  at the {ILC}},
\newblock Journal of Instrumentation
\newblock {\bf  2} (2007), P09001.

\bibitem{Jad97}
S. Jadach, W. Placzek, E. Richter-Was, B. Ward, and Z. Was,
\newblock {\it Upgrade of the {Monte Carlo} program {BHLUMI} for {Bhabha}
  scattering att low angles to version 4.04},
\newblock Computer Physics Communications
\newblock {\bf  102} (1997), 229.

\bibitem{IDR11}
{ILC Collaboration}
\newblock {\it {International Linear Collider - A technical progress report}}
\newblock Technical Report ILC-REPORT-2011-30, {International Linear Collider},
  2011 ({ISBN}: 978-3-935702-56-0).

\bibitem{Stahl05a}
A. Stahl
\newblock {\it Diagnostics of colliding bunches from pair production and
  beamstrahlung at the {IP}}
\newblock Technical Report LC-DET-2005-003, {DESY}, 2005
  (http://www-flc.desy.de/lcnotes/).

\bibitem{Grah08}
C. Grah and A. Sapronov,
\newblock {\it Beam parameter determination using beamstrahlung photons and
  incoherent pairs},
\newblock Journal of Instrumentation
\newblock {\bf  3} (2008), P10004.

\bibitem{Sad08}
I. Sadeh
\newblock {\it Luminosity measurement at the {International Linear Collider}}
\newblock Master's thesis, Tel Aviv University, 2008.

\bibitem{Agu11}
J. Aguilar et~al.,
\newblock {\it Luminometer for the future international collider - simulation
  and beam test},
\newblock in Technology and Instrumentation in Particle Physics - TIPP 2011,
\newblock 2011
\newblock (preprint available at http://arxiv.org/abs/1111.5199).

\bibitem{Schw11}
R. Schwartz,
\newblock {\it {LumiCal} performance in the {CLIC} environment - progress
  report},
\newblock in Proceedings of the 18th FCAL Workshop in Predeal, Romania,
\newblock 2011, p.~84.

\bibitem{Abr07}
H. Abramowitz, R. Ingbir, S. Kananov, and A. Levy
\newblock {\it {A luminosity detector for the International Linear Collider}}
\newblock Technical Report LC-DET-2007-006, {DESY}, 2007
  (http://www-flc.desy.de/lcnotes/).

\bibitem{ILD10}
{ILD Concept Group}
\newblock {\it {The International Large Detector - Letter of Intent}}
\newblock Technical Report DESY 2009/87 - Fermilab PUB-09-682-E - KEK Report
  2009-6, {DESY / KEK / Fermilab}, 2010 ({ISBN}: 978-3-935702-42-3,
  http://ilcild.org/documents/ild-letter-of-intent/).

\bibitem{Abr10}
H. Abramowicz et~al.,
\newblock {\it Forward instrumentation for {ILC} detectors},
\newblock Journal of Instrumentation
\newblock {\bf  5} (2010), P12002.

\bibitem{Mila11}
M. Pandurovi\'{c},
\newblock {\it Fon u merenju luminoznosti i razvoj metode za identifikaciju
  $b$-kvarka u eksperimentima ILC i H1},
\newblock PhD thesis, Univerzitet u Beogradu, 2011.

\bibitem{Arb96}
A. Arbuzov et~al.,
\newblock {\it The present theoretical error on the {Bhabha} scattering cross
  section in the luminometry region at {LEP}},
\newblock Physics Letters B
\newblock {\bf  383} (1996), 238 -- 242.

\bibitem{Mila13}
M. Pandurovi\'c and I. Bo\v{z}ovi{\'c}-Jelisav\v{c}i{\'c}
\newblock {\it Physics background at {ILC} at {500~GeV} and {1~TeV}}
\newblock Technical Report under review PREL-LC-DET-2012-011, {DESY}, 2013
  (http://www-flc.desy.de/lcnotes/; Preprint available at arXiv:1301.2494).

\bibitem{Poz07}
V. Pozdnyakov,
\newblock {\it Two-photon interactions at {LEP}},
\newblock Physics of Particles and Nuclei Letters
\newblock {\bf  4} (2007), 289--303.

\bibitem{RDR07}
(T. Behnke, C. Damerell, J. Jaros, and A. Miyamoto, eds.),
\newblock {\it {International Linear Collider - Reference Design Report}},
\newblock Vol.~4,
\newblock {ILC} Global Design Effort and {World} Wide Study,
\newblock 2007.

\end{thebibliography}

\end{document}